\documentclass[12pt]{iopart}

\usepackage{iopams}
\usepackage{epsfig}
\usepackage{graphicx}
\usepackage{cite}
\usepackage{amssymb}
\usepackage{array}

\newcommand{\fd}[2]{\frac{d #1}{d #2}}

\newcommand{\pd}[2]{\frac{\partial}{\partial #2} #1}

\begin{document}

\title{Mutual information via thermodynamics: Three different approaches}

\author{Yitzhak Peleg$^{1}$, Hadar Efraim$^{1}$, Ori Shental$^{2}$ and Ido Kanter$^{1}$}

\address{$^1$  Minerva Center and Department of Physics, Bar-Ilan University, Ramat-Gan 52900, Israel}
\address{$^2$ Center for Magnetic Recording Research (CMRR), University of
California, San Diego (UCSD), 9500 Gilman Drive, La Jolla, CA 92093,
USA}


\begin{abstract}
Three different approaches to derive mutual information
via thermodynamics are presented where the temperature-dependent
energy is given by: (a) $\beta \mathcal{E} = -\ln[P(X,Y)]$, (b)
$\beta \mathcal{E} =-\ln[P(Y|X)]$ or (c) $\beta \mathcal{E}
=-\ln[P(X|Y)]$. All approaches require the extension of the
traditional physical framework and the modification of the 2nd law
of thermodynamics. A realization of a physical system with an
effective temperature-dependent Hamiltonian is discussed followed by
a suggestion of a physical information-heat engine.
\end{abstract}

\maketitle

\section{Introduction}
\begin{figure}[t!]
\label{communication channel}
\begin{center}
\includegraphics[scale=0.7]{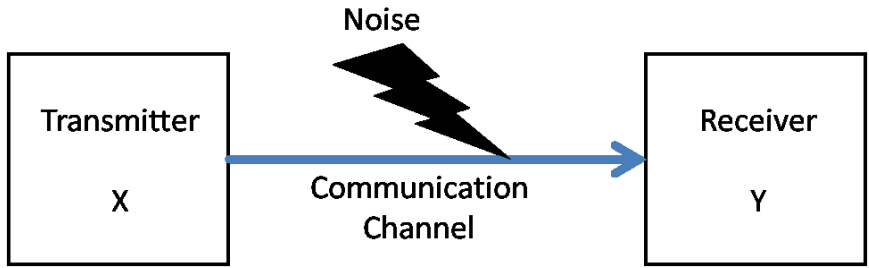}
\caption{A schematic communication channel.}
\end{center}
\end{figure}

The generic problem in information processing is the transmission of
information over a noisy communication channel \cite{Cover,Blahut,Gallager}.
The transmission can be mathematically described by two random variables
$X$ and $Y$ representing the desired information and its noisy
replica, respectively.
A schematic figure of a communication channel is depicted in Fig. \ref{communication channel} .
The basic properties of a communication system are: $P(X)$ which is the
probability of transmitting a symbol $X$ taken from the input
alphabet, and $P(Y|X)$ which stands for the probability of receiving a symbol
$Y$ (taken from the output alphabet) following the transmission of
a symbol $X$. Noisy transmission can occur either via space from one
geographical point to another, as happens in communications, or in
time, for example, when sequentially writing and reading files from
a hard disk in the computer.

Mutual information, $I(X;Y)$, is a principle quantity in information theory
which quantifies the amount of information in common between two
random variables. It is used to upper bound the attainable rate of information transferred across a channel.
A basic definition of the mutual information is
\begin{equation}
\label{Def. Mutual Infromation} I(X;Y) \equiv H(X)-H(X|Y) =
H(Y)-H(Y|X),
\end{equation}
where $H(\cdot)$ is the Shannon's information entropy (in Nats)
\cite{Shannon}. The mutual information measures the amount of
uncertainty in a random variable, indicating how easily data can be
losslessly compressed. Hence knowing $Y$, we can save an average of
$I(X;Y)$ bits in encoding $X$ compared to not knowing $Y$
\cite{Cover,Nisimori}.

A fundamental link between information theory and thermodynamics was
first established five decades ago by Jaynes \cite{Jaynes}. However,
his work did not include an explicit relation between mutual
information and thermodynamics.

Recently, it has been proven
\cite{Shannon_meets_Carnot1,Shental&Kanter-epl} that the mutual
information can be reformulated as a consequence of the laws of
thermodynamic, where the corollary was exemplified for the Gaussian
noisy channel and for the binary symmetric channel. The modeling of
the communication channels as a thermal system required the
generalization of thermodynamics to include T-dependent Hamiltonians
and the generalization of the second law of thermodynamic which was
proved to have the following form
\begin{equation}
\label{generalized second law}
dQ = TdS+ \left< \frac{d\mathcal{E}}{dT} \right> dT,
\end{equation}
where $\langle \cdot \rangle$ denotes averaging over the standard Boltzmann distribution.

In communication channel the goal is to estimate the transmitted symbol $X$ from the received symbol $Y$ (Fig. \ref{communication channel}), hence the main quantity of interest is $P(X|Y)$. A physical system with equivalent properties as the communication channel has to obey the following
\begin{equation}
\label{Basic Eq}
\frac{e^{-\beta\mathcal{E}}}{\mathcal{Z}}
=P(X|Y)=\frac{P(X)P(Y|X)}{P(Y)} .
\end{equation}

This new bridge between mutual information and thermodynamics
requires the extension of the traditional physical framework and the
following two questions are at the center of the first part of our
work. The first one is whether the mapping between mutual
information and thermodynamics as well as the physical energy
governing a given communication channel is uniquely defined. In case
the energy function is not uniquely defined, the question is whether
the required extension of the physical framework is a necessary
ingredient, or there is a physical way to express the mutual
information using the traditional physical framework without
altering the second law of thermodynamics.

The answers to the above questions are that the mapping between mutual
information and thermodynamics as well as the physical energy
governing a given communication channel is not uniquely defined and
requires the extension of the physical framework.
In the following we present three primary approaches, followed by a discussion of a possible physical system with an effective T-dependent Hamiltonian and a possible realization of an information-heat engine. Details
of the derivations are left for \ref{Derivations}, whereas \ref{Applications}
exemplifies the calculation of the mutual information of a few
archetypal communication channels via thermodynamics.

\section{1st approach - Boltzmann factor $\propto P(X,Y)$}
This approach takes the joint probability $P(X,Y)$ to be the Boltzmann factor and defines a T-dependent energy,
\begin{equation}\label{energy definition p(x,y)}
\mathcal{E} = -\frac{1}{\beta}\ln[P(X,Y)].
\end{equation}

This form of the energy is a naive physical energy definition, since
it adequately describes a physical system consisting of two degrees
of freedom, $X$ and $Y$, in a contact with a macroscopic heat
reservoir. At equilibrium the expectation properties of $X$ and $Y$
are determined following the partition function  $\mathcal{Z}=
\sum_\mathcal{E} \exp  (-\beta \mathcal{E} )$ \cite{Reif,Reichl}.
Using the T-dependent Hamiltonian (\ref{energy definition p(x,y)}),
to describe a communication channel (e.g.
\cite{Shannon_meets_Carnot1}, Eq. 44) enforces the generalization of
the second law of thermodynamics (\ref{generalized second law}), and the mutual
information takes the following form
\cite{Shannon_meets_Carnot1,Shental&Kanter-epl}
\begin{equation}
\label{Eq. Mutual information Shental&kanter}
 I(X;Y) =
-\mathbb{E}_{Y;\beta}
\left\{
\gamma U  \left.\right|_{\gamma=0}^{\gamma=\beta}-
\int_0^\beta \left (U + \gamma \left< \pd{\mathcal{E}}{\gamma} \right>\right)d\gamma
\right\}
\end{equation}
where $\mathbb{E}_{Y;\beta} \{\cdot\}$ denotes
expectation of the random object within the bracket with respect to
the subscript random variable $Y$, and for a given temperature $\beta$.

\section{2nd approach - Boltzmann factor $\propto P(Y|X)$}
This approach refers to $P(Y|X)$ as the Boltzmann factor \cite{Tishby} and  the resulted 
energy is
\begin{equation}\label{energy definition p(y|x)}
\mathcal{E} = -\frac{1}{\beta}\ln[P(Y|X)].
\end{equation}

The definition of the energy, (\ref{energy definition p(y|x)}), is
based on the interpretation of the prior probability of the inputs
of the channel, $P(X)$, as the degeneracy of the energy level
$\mathcal{E}$ \cite{Tishby}. This approach was  recently adopted also by \cite{Merhav}. It depicts a scenario of communication
channels where the output, $Y$, is estimated by the input, $X$.
Since the degeneracy of the input of the channel can be designed
arbitrary, the degeneracy of the physical energy function
(\ref{energy definition p(y|x)}) may decrease while the energy
increases, in contrast to physical systems.  The emission of heat to
the reservoir decreases the energy and increases the entropy. Hence,
both terms of the free energy identity, $F=U-TS$, decrease and the
system is unstable to thermal fluctuations. This situation demands a
modification of both the free energy and the second law of
thermodynamics (\ref{free_energy_with dkl},\ref{second law_with
dkl}).

The mutual information for energy (\ref{energy definition p(y|x)})
is given by
\begin{equation}\label{mutual p(y|x)}I(X;Y)=-\mathbb{E}_{Y;\beta} \left\{\beta(U-F)\right\}.\end{equation}
where the free energy (as implicitly suggested in
\cite{Shental&Kanter-epl}) and the second law are effectively
modified to be
\begin{equation}\label{free_energy_with dkl}F=U + T D_{kl}(P(X|Y)||P(X))\end{equation}
and
\begin{equation}\label{second law_with dkl}dQ=-Td D_{kl}(P(X|Y)||P(X)),\end{equation}
respectively. For a clarification, $F$ is the free energy,
$D_{kl}(\cdot)$ denotes the Kullback-Leibler
divergence\cite{Cover} and the Boltzmann constant is arbitrarily
taken to be unity. The derivation of (\ref{mutual p(y|x)},\ref{free_energy_with dkl},\ref{second law_with dkl})
is detailed in \ref{Derivations}. Note that when $P(X)$ is uniformly distributed, the conventional
identity of the free energy, $F=U-TS$, and the second thermodynamic
law, $dQ=TdS$, are restored.

\section{3rd approach - Boltzmann factor $\propto P(X|Y)$}
We propose a new approach, where we define the Boltzmann factor to be $P(X|Y)$. As a result, the T-dependent energy is
\begin{equation}\label{energy definition p(x|y)}
\mathcal{E} = - \frac{1}{\beta}\ln[P(X|Y)].
\end{equation}

The 3rd approach describes a typical communication system where the
input, $X$, is estimated by the output, $Y$. Nevertheless, in this
kind of energy functions, the partition function is normalized to
$\mathcal{Z}=1$, independent of the temperature. Yet also this
approach requires the generalization of the second law
(\ref{generalized second law}), however, the mutual information has
a simple form of the internal energy only
\begin{equation}
\label {new formation of I} I(X;Y)  =
-\mathbb{E}_{Y;\beta}\left\{\gamma U
|_{\gamma=0}^{\gamma=\beta}\right\}
.
\end{equation}

Note that the proposed approach encompasses the other two approaches
\cite{Shannon_meets_Carnot1,Tishby,Shental&Kanter-epl}. On the one
hand, the mutual information (\ref{new formation of I}) can easily
be deduced from the 2nd approach using $F=0$ in (\ref{mutual p(y|x)}). 
On the other
hand, the energy function (\ref{energy definition p(x|y)})
explicitly indicates that the second term of Eq. (\ref{Eq. Mutual
information Shental&kanter}) is identically zero. A comprehensive
derivation of the 3rd approach is exhibited in \ref{Derivations}.

A synopsis of a comparison between the three approaches is depicted
in Table \ref{tab1}. 

\begin{table}
\tiny
\begin{tabular}{|c || c | c | c |}
\hline
&&&\\
$\beta \mathcal{E}$ &   $I(X;Y)$    &   $\mathcal{F}$   &   $2^{nd}$ law
\\&&&\\ \hline
&&&\\
$-\ln[P(X,Y)]$      &
$
\left.-\left\{\mathbb{E}_{Y;\gamma}
\left( \gamma U - \int_0^\gamma U + \gamma \left< \pd{\mathcal{E}}{\gamma} \right>_{X|Y;\gamma}\right) d\gamma \right\}
\right|_{\gamma=0}^{\gamma=\beta}$ & $U-TS$ &
$dQ=T dS + \mathbb{E}\bigg\{\fd{\mathcal{E}}{T}\bigg\}dT$
\\&&&\\ \hline
&&&\\
$-\ln[P(Y|X)]$      &   $\mathbb{E}_{Y} \left\{\beta(F-U)\right\}$
    &   $U-TD_{kl}(P(X|Y)||P(X))$   &   $d Q=Td D_{kl}$
\\&&&\\ \hline
&&&\\
$-\ln[P(X|Y)]$      &
        $\left\{ \mathbb{E}_{Y;\gamma}\left(-\gamma U\right) \right\}  \left|_{\gamma=0}^{\gamma=\beta}\right.$    &   $U-TS=0$    &$dQ=T dS + \mathbb{E}\bigg\{\fd{\mathcal{E}}{T}\bigg\}dT$
\\&&&\\ \hline
\end{tabular}
\normalsize \caption{A comparison between the three approaches to
connect the mutual information via thermodynamics. Each approach
requires the extension of the traditional physical framework and
yields modified definitions  for the free energy and/or for the
second law of thermodynamics.\label{tab1}}
\end{table}

\section{Physical information-heat engine}

The extension of the physical framework to include T-dependent
Hamiltonians and the generalized second law of thermodynamics might
also refresh our viewpoint on traditional physical systems.

A prototypical physical system governed by an effective T-dependent
Hamiltonian is a spring where the spring constant is a function of
the temperature, $K=K(T)$ \cite{spring}. The energy of the spring is
\begin{equation}
\label{spring_energy}
\mathcal{E}=\frac{1}{2} K(T)z^2
\end{equation}
where $z$ denotes the extension of the spring from a reference
position with the lack of force on the spring.
Note that the common scenario is that the free energy is an explicit function
of the temperature. However, in our case, the (effective)
\textbf{Hamiltonian} is a function of the temperature. This dependence calls
for an explanation, since the fundamental potentials (gravitation,
electromagnetic etc.) governing the known physical laws are
independent of the temperature. The solution of this mystery, a T-dependent Hamiltonian, is that
the spring is represented by one macroscopic degree of freedom and
its property is a consequential of a coarse grained over the
microscopic many degrees of freedoms and the nonlinear forces among
them.

\begin{figure}[t]
\begin{center}
\includegraphics[scale=0.7,angle=90]{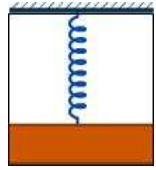}
\caption{A spring, with a temperature dependent spring constant,
$K(T)$, connected to a mass $M$ hangs in a container which is
vacuumed.
\label{spring}
}
\end{center}
\end{figure}

\begin{figure}[t]
\begin{center}
\includegraphics[scale=0.25]{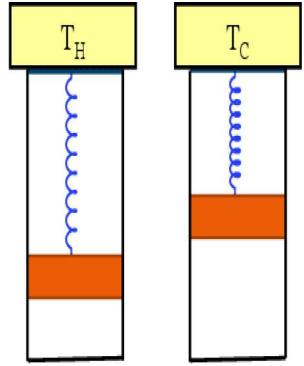}
\caption{A container in a thermal contact with a heat-reservoir at high
temperature, $T_{H}$, (left panel) and for a heat-reservoir at cold
temperature, $T_{C}$, (right panel). We assume that the spring
constant monotonically decreases with the temperature, hence the
extension of the spring at $T_{H}$ is greater than for
$T_{C}$.
\label{basic_two}
}
\end{center}
\end{figure}

A mass, $M$, is connected to one end of the spring and the spring
with the connected mass is hanged in a container which is vacuumed
(Fig. \ref{spring}). The container is connected to a heat reservoir
at a temperature $T$ and for the simplicity of the following
discussion  we assume that the spring constant monotonically
decreases with the temperature. The equilibrium situation at two
different temperatures, $T_{H}>T_{C}$ is depicted in Fig.
\ref{basic_two}. We turn now to describe a possible
information-heat engine based on such T-dependent Hamiltonian.

\begin{figure}[t]
\begin{center}
\includegraphics[scale=0.25]{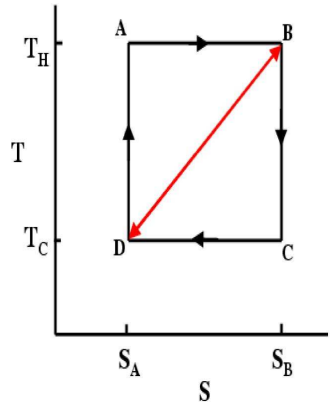}
\caption{ A Carnot cycle acting as a heat engine, illustrated on a
temperature-entropy diagram. The vertical axis is temperature, the horizontal axis is
entropy. The cycle takes place between a hot
reservoir at temperature $T_{H}$ and a cold reservoir at temperature
$T_{C}$.
\label{carnot_cycle}
}
\end{center}
\end{figure}

A Carnot cycle acting as a heat engine, is illustrated by the
black-cycle in the temperature-entropy diagram in Fig.
\ref{carnot_cycle}. The Carnot cycle consists of $4$-steps,
alternating isothermal and adiabatic processes \cite{Reif,Reichl}.
The cycle of the information-heat engine consists of two steps only
and is illustrated by the red lines in Fig. \ref{carnot_cycle}. The
first step, from D to B, describes a quasi-static process where the
temperature increases from $T_{C}$ to $T_{H}$ and both the
temperature and the entropy increase, since the Hamiltonian is an
explicit function of the temperature. In the reversed process, from
B to D, the temperature decreases in a quasi-static manner back to
$T_{C}$ and the cycle is completed. No work is done in the entire
cycle, D-B-D, since the container is vacuumed, and mathematically
the area formed by the cycle D-B-D in the $(S,T)$ plane is zero.

The heat absorbed/emmited by the C/H reservoirs is responsible for
the following two main changes of the system (spring+mass) placed in
the container: (a) The kinetic energy of the microscopic degrees of
freedom is modified. (b) The Hamiltonian of the spring is
modified via the T-dependent spring constant.
This process was named as "channel work" in
\cite{Shental&Kanter-epl,Thermodynamic_derivation_of_the_mutual_information},
since it reduces the effective heat contributing to the change in
the entropy and it resembles work. However, no actual work is done.  Note
that in principle at equilibrium the macroscopic mass, M, oscillates
as an harmonic oscillator too, since each degree of freedom has on
the average a kinetic energy equals to $K_{B}T/2$, however, these
microscopic vibrations are neglected.

The information-heat engine depicted in Fig. \ref{carnot_cycle}
describes a way to generate bits in a way resembling a traditional
heat engine, but with the lack of work.  The height of the mass M
represents the generated bit: the cold position represents "0"
whereas the hot position represents "1". A generation of a sequence
of bits can be done by using a predetermined protocol indicating the
frequency (bandwidth) for the generation of bits. For instance, in
the event that the current bit is "0" and the successor bit is "0"
too, the contact to the cold reservoir remains, but in case of a
successor "1", the container is brought to a contact with the hot
reservoir.

The proposed information-heat engine describes a way to generate the
information, a sequence of bits, in a 2-steps cycle and with the
lack of work. The generation of the communication channel requires a
fundamental physical mechanism to transmit the bits and with minimal
work in order to enhance the efficiency of the process . All such
mechanisms have to "read" and to estimate the height of the mass in
the container. Note that the framework of noisy  communication
channel enables a distortion of the information, however, the
encoder represents a noise-free process where the noise is added
during the transmission only. Hence, the information-heat engine has
a lack of inherent noise. There are many possible mechanisms to
estimate the height of the mass using, for instance,
reflected/transmitted photons from the mass/lack-of-mass at a given
height, however, it is beyond the scope of our work.

In the above, we presented a possible mechanism which resembles the Carnot engine, but
with the lack of work.
There are many alternative physical ways to
generate an information-heat engine. For instance, using a material which undergoes a
ferromagnetic/paramagnetic transition in between $T_{C}/T_{H}$.
However, the essence of such an information-heat engine is a T-dependent Hamiltonian.

\appendix
\section{Derivations of the 2nd and 3rd approaches}
\label{Derivations}
\subsection{Derivation of the 2nd approach}
The minimal mutual information for a given expected distortion,
$\mathbb{E}_{X,Y} \{d(Y,X)\}$, can be found by minimizing the functional
$\mathcal{F}(P(X|Y))=I(X;Y)+\beta \ \mathbb{E}_{X,Y} \{d(Y,X)\}$ over all normalized distributions $P(X|Y)$ \cite{Tishby}.
The  solution of the variational problem is the normalized probability
\begin{equation}
\label{P(X|Y)}
    P(X|Y) =\frac{P(X)}{Z(Y,\beta)}e^{-\beta d(Y,X)},
\end{equation}
where $\ln{Z(Y,\beta)} =\frac{\lambda(Y)}{P(Y)}$. $\lambda(Y)$ and $\beta$ are the Lagrange multipliers of the normalization and the expected distortion constraints, respectively.
Moreover, $\beta$ is positive and satisfies\cite{Tishby}
\begin{equation}
\label{Lagrange multiplier beta}
    \beta  = -\frac{\delta I(X;Y)}{\delta \ \mathbb{E}_{X,Y} \{d(Y,X)\}}.
\end{equation}

In order to satisfy the energy definition (\ref{energy definition p(y|x)}) and using the Bayes' law,
a comparison of (\ref{P(X|Y)}) with the Boltzmann distribution law yields the following mapping:
$d(Y,X) \to \mathcal{E}$, $\beta \to 1/T$ , $P(X)$ is used as the degeneracy of the energy level $d(Y,X)$ and $Z(Y,\beta)$ is the partition function for a given $Y$.

The internal energy of the system, $U$, is the expectation value of the energy, $d(Y,X)$.
By equating the Lagrange multiplier, $\beta$ (\ref{Lagrange multiplier beta}) to the second law of thermodynamics, $\beta=\frac{dS}{dU}$, it is easy to see that this system obeys the following mapping
\begin{equation}
\label{mapping S to I}
S \to -I.
\end{equation}

A verification of (\ref{mapping S to I}) can be observed using the relation
\begin{equation}
\label{Free Energy with I}
F=T\mathcal{F}=U+T I,
\end{equation}
followed by comparing the free energy identity, $F=U-TS$, to $\mathcal{F}$ (See eq. 12 in \cite{Tishby}).

Substituting $I=\mathbb{E}_{Y;\beta}
\left\{D_{kl}(P(X|Y)||P(X))\right\}$ \cite{Tishby} into Eqs.
(\ref{Lagrange multiplier beta}, \ref{Free Energy with I}) and based
on the first law of thermodynamics with the lack of work, $dQ=dU$,
we obtain Eqs. (\ref{free_energy_with dkl},\ref{second law_with
dkl}).

\subsection{Derivation of the 3rd approach}
The definitions of the marginal and conditional entropies,
consisting the mutual information (\ref{Def. Mutual Infromation}),
are $ H(X) \equiv -\sum_X P(X) \ln{P(X)} $ and $ H(X|Y)  \equiv
-\sum_{X,Y} P(X,Y) \ln{P(X|Y)}$, respectively. Note that when $X$ and
$Y$ are independent random variables, $H(X|Y)$ becomes $H(X)$.

We introduce a new variable $\beta$ which represents the noise in the channel and has the following properties
\begin{eqnarray}
\label{Def. beta} P(X,Y; \beta=0) =  P(X)P(Y ;
\beta=0)
\nonumber \\
P(X|Y; \beta=0) =  P(X),
\end{eqnarray}
where $P(Y ; \beta)$ is the probability of receiving $Y$ for a given noise
$\beta$.
As a result, the conditional entropy becomes an explicit function of $\beta$. Using Bayes' law and defining
\begin{equation}
\label{S_definition}
S(X|Y;\beta)= - \sum_X P(X|Y;\beta) \ln P(X|Y;\beta),
\end{equation}
we can write the entropies as,
\begin{eqnarray}
   H(X) = S(X|Y;\beta=0)
\\
\label{entopies via S}
   H(X|Y ;\beta)  = \mathbb{E}_{Y;\beta}\left\{ S(X|Y;\beta)\right\}
.
\end{eqnarray}
Note that a noiseless channel is represented by the limit ${\beta\to \infty}$.
Taking into account that $H(X)$ is independent of $Y$,
we can write,
\begin{equation}
\label{Eq. Hx as Hxy0} H(X)=\mathbb{E}_{Y;\beta}\left\{ S(X|Y;\beta=0)\right\}.
\end{equation}
Substituting Eqs. (\ref{entopies via S},\ref{Eq. Hx as Hxy0})  into Eq.
(\ref{Def. Mutual Infromation}), we achieve a new form of the mutual
information
\begin{equation}
\label{Eq. Mutual Information} I(X;Y)  =
-\mathbb{E}_{Y;\beta}
\left\{S(X|Y;\gamma)
\left|_{\gamma=0}^{\gamma=\beta}\right.\right\}.
\end{equation}

Following (\ref{energy
definition p(x|y)}), it is clear that the partition function of the equivalent thermodynamic system is $\mathcal{Z}=1 $, hence Eq. (\ref{S_definition}) is
\begin{equation}
\label {Eq. diffrent energy - S} S(X|Y;\gamma) =  -\sum_X
e^{\left(-\gamma \mathcal{E}\right)}  \left(-\gamma
\mathcal{E}\right)=\gamma U,
\end{equation}
where $U$ is the thermodynamic average of $\mathcal{E}$, divided
by the partition function, or in other words, the internal energy.
The free energy obeys $F=-\frac{1}{\beta}\ln{Z}=U-\frac{S}{\beta}=0$, hence $S=\beta U$.
Substituting (\ref{Eq. diffrent energy - S}) into (\ref{Eq. Mutual
Information}), we finally receive a much simpler thermodynamic form
of the mutual information, as the difference of the internal energies of the system,
\begin{equation}
\label {Eq. diffrent energy - I} I(X;Y)  =
-\mathbb{E}_{Y;\beta}
\left\{\gamma U
\left|_{\gamma=0}^{\gamma=\beta}\right.
\right\}
.
\end{equation}

\section{Applications of the 3rd approach}
\label{Applications}
The new description of the mutual information (\ref{new formation of I}, \ref{Eq. diffrent
energy - I})
is exemplified over several archetypal communication
channels. The sketch of the calculations for the Gaussian channel
with Gaussian input, the Gaussian channel with Bernoulli-1/2 input
and finally the binary symmetric channel with a biased input (Biased
BSC) are presented. For the examples we shall use the following
notations: $P(X)\equiv P(X=x), P(Y)\equiv P(Y=y)$.

\subsection{Gaussian channel with $\mathcal{N}(0,1)$ input}
The input and the a-posteriori probabilities of this channel are
\begin{eqnarray}
P(X) = \mathcal{N}(0,1)
\nonumber\\
P(Y|X) = \mathcal{N}(0,\frac{1}{\beta}).
\end{eqnarray}
Hence, the energy, according to Bayes' law and (\ref{energy definition p(x|y)}) is
\begin{equation}
\mathcal{E}=\frac{x^2}{2} \left (\frac{1+\beta}{\beta} \right )-x
y-\frac{\ln (\beta +1)}{2 \beta}+\frac{y^2}{2} \left
(\frac{\beta}{1+\beta} \right )+ \frac{\ln (2 \pi )}{2 \beta}.
\end{equation}
Using  (\ref{Eq. diffrent energy - I}) one can easily find the formula for
the Shannon capacity \cite{Shannon},
\begin{equation}
\label{Gauss mutual information} I(X;Y)= \frac{1}{2} \ln (1+\beta),
\end{equation}
which is identical to the mutual information derived from the 1st approach, eqs.  (\ref{energy definition p(x,y)},\ref{Eq. Mutual information
Shental&kanter})
\cite{Shental&Kanter-epl}.

\subsection{Gaussian channel with Bernoulli-1/2 input}
This case is characterized by equiprobable binary inputs a-posteriori probabilities as following
\begin{eqnarray}
P(X=1) = P(X=-1) =1/2
\nonumber \\
P(Y|X) = \mathcal{N}(0,\frac{1}{\beta}).
\end{eqnarray}
Implementing the Baye's
law, while dropping the elements which are independent of $x$,
yields a simple expression for the energy (\ref{energy definition
p(x|y)}),
\begin{equation}  \mathcal{E}=-xy -\frac{\ln(2 \cosh (y
\beta))}{\beta},
\end{equation}
which eventually gives us the known Shannon-theoretic result \cite{Blahut},
\begin{equation}
\label{Bernoulli first term x|y}
I(X;Y)=
\nonumber\\
\beta  - \frac{1}{\sqrt{2 \pi}} \int _{-\infty}^{\infty}
exp\left(\frac{-y^2}{2} \right)\log \cosh \left(\beta-\sqrt{\beta} y\right) dy.
\end{equation}

\subsection{Biased binary symmetric channel}
In this case the prior distribution of the biased input and the probability for
a symbol to flip during the transmission are denoted as
\begin{eqnarray} P(X=-1)=p \nonumber\\ P(X=1)=1-p   \end{eqnarray} and \begin{equation}P(Y= \pm
1|X= \mp 1) = \delta\end{equation} respectively.
Hence, the probabilities are defined as following,
\begin{eqnarray}
P(X)    =  p^\frac{1-x}{2} \left(1-p\right)^\frac{1+x}{2} &\qquad  x\in\left\{-1,1\right\} \nonumber \\
P(Y|X)  =  \delta^\frac{1-x y}{2} \left(1-\delta\right)^\frac{1+x
y}{2}   &\qquad   y\in\left\{-1,1\right\}.
\end{eqnarray}
The energy (\ref{energy definition p(x|y)}) is now given by
\begin{eqnarray}
\mathcal{E} = & -\frac{x y}{2}+\frac{\ln
   ((1-\delta)\delta)}{2 \beta}-\frac{x \ln \frac{1-p}{p}}{2 \beta}
   - \frac{\ln((1-p)p)}{2 \beta} + \nonumber \\ & \left(\frac{1+y}{2}\right) \ln(1-p-\delta+2 p
   \delta) + \left(\frac{1-y}{2}\right) \ln(p+\delta-2 p \delta).
\end{eqnarray}
where the inverse temperature, $\beta$, is defined as
\begin{equation}
\label{beta delta}
\beta=\ln \left (\frac{1-\delta}{\delta}\right ).
\end{equation}
Applying (\ref{Eq. diffrent energy - I}), we finally receive the
mutual information,
\begin{eqnarray}
\label{Biased BSC mutual information} I(X;Y) = &
 \delta \ln (\delta ) +\left(1-\delta\right)  \ln \left(1-\delta\right) \nonumber \\ & -(p+\delta-2 \delta  p ) \ln (p+\delta-2
\delta  p )\nonumber \\ & -(1-p-\delta+2\delta  p) \ln
(1-p-\delta+2\delta  p).
\end{eqnarray}
For $p=\frac{1}{2}$ the mutual information for the BSC is restored
\cite{Thermodynamic_derivation_of_the_mutual_information}
$$ I(X;Y)= \delta \ln(\delta) + (1-\delta)\ln(1-\delta) + \ln2 .$$

Note that $\beta$ is a function of the noise $(\delta)$, solely, i.e. it is not effected by the nature of the input, $P(X)$.
\\\\


\begin{thebibliography}{99}
\bibitem{Cover}
Cover T M and Thomas J A 1991 \emph{Elements of Information Theory}
(New York: Wiley)

\bibitem{Blahut}
Blahut R E 1987 \emph{Principles and Practice of Information Theory}
(Reading, MA: Addison-Wesley)

\bibitem{Gallager}
Gallager R G 1968 \emph{Information Theory and Reliable
Communication} (New York: Wiley)

\bibitem{Shannon}
Shannon C E 1948 Bell Syst. Tech. J. \textbf{27}

\bibitem{Nisimori}
Nishimori H 2001 \emph{Statistical Physics of Spin Glasses and Information Processing} (Oxford University Press, Oxford, UK)

\bibitem{Jaynes}
Jaynes E T 1957 Phys. Rev. \textbf{106} 620-630


\bibitem{Shannon_meets_Carnot1}
Shental O and Kanter I 2008 arXiv:0806.3133v1

\bibitem{Shental&Kanter-epl}
Shental O and Kanter I 2009, Europhysics Let. \textbf{85}

\bibitem{Reif}
Reif F 1965 \emph{Fundamental of Statistical and Thermal Physics} (McGrew-Hill, New-York)

\bibitem{Reichl}
Reichl L E 1998 \emph{A Modern Course in Statistical Physics}
(Wiley, New-York)

\bibitem{Tishby}
Tishby N, Pereira F C and Bialek W 1999, 37th Allerton Conf. on Commun. Control and Computing

\bibitem{Merhav}
Merhav N 2008 IEEE Trans. Inform. Theory \textbf{54}

\bibitem{spring} Kohl W H 1967 \emph{Handbook of Materials and Techniques for Vacuum Devices}
(Reinhold Publishing Corp., New York)

\bibitem{Thermodynamic_derivation_of_the_mutual_information}
Kanter I, Shental O, Efraim H and Yacov N 2008 J. Phys. A: Math.
Theor. \textbf{41}

\end{thebibliography}
\end{document}